\documentstyle[preprint,prd,aps,epsfig]{revtex}
\def\ee{$e^+e^-$}

\def\alr{A_{LR}}
\def\alrraw{A^{obs}_{LR}}
\def\qlr{A_Q^{obs}}
\def\qlrraw{A_Q^{obs}}
\def\qr{Q}
\def\tqr{\tilde{Q}}
\def\swein{\sin^2\theta_W^{\rm eff}}
\def\pole{{\cal P}_e}
\def\absp{\vert{\cal P}_e\vert}
\def\z0{$Z$}
\def\xl{\sigma_L}
\def\xr{\sigma_R}

\begin{document}
\draft
\preprint{\vbox{\hsize=220pt\noindent  SLAC--PUB--7069, COLO--HEP-367 \\
September 1996 \\ (E/T)}}
\title{First Measurement of the Left-Right Charge Asymmetry\\
       in Hadronic $Z$ Boson Decays and a New Determination\\
               of ${\swein}$$^\dagger$}
\author{The SLD Collaboration$^*$}
\address{Stanford Linear Accelerator Center\\
         Stanford University, Stanford, California, 94309\\}
\maketitle
%
\begin{abstract}
We present the first measurement of the left-right charge asymmetry 
$A_Q^{obs}$ in hadronic $Z$ boson decays. This was performed at E$_{\rm cm}$ = 
91.27~GeV with the SLD at the SLAC Linear Collider with a polarized 
electron beam. Using 89838 events, we obtain $A_Q^{obs}=0.225 \pm 0.056~\pm 
0.019$ which leads to a measurement of the electron left-right asymmetry
parameter, $A_e = 0.162 \pm 0.041~\pm 0.014~$, and $\swein = 0.2297\pm 0.0052
\pm 0.0018$. Also, the $A_Q^{obs}$ measurement combined with the left-right 
cross section asymmetry determines $A_e$ independent of the value of the 
electron-beam polarization.
\end{abstract}
%
%
\vskip 0.2in
\begin{center} {\rm Submitted to {\em Physical Review Letters}} \end
{center}
\vskip 0.5in
\vbox{\footnotesize\renewcommand{\baselinestretch}{1}\noindent
$^\dagger$This work was supported in part by Department of Energy contract 
DE-AC03-76SF00515} 



\bigskip
\begin{center}
\bigskip
%
%
%
  \def\iADEL{$^{(1)}$}
  \def\iBOL{$^{(2)}$}
  \def\iBU{$^{(3)}$}
  \def\iBRUN{$^{(4)}$}
  \def\iUCSB{$^{(5)}$}
  \def\iUCSC{$^{(6)}$}
  \def\iCIN{$^{(7)}$}
  \def\iCSU{$^{(8)}$}
  \def\iCOLO{$^{(9)}$}
  \def\iCOL{$^{(10)}$}
  \def\iFER{$^{(11)}$}
  \def\iFRA{$^{(12)}$}
  \def\iILL{$^{(13)}$}
  \def\iLBL{$^{(14)}$}
  \def\iMIT{$^{(15)}$}
  \def\iMASS{$^{(16)}$}
  \def\iMISS{$^{(17)}$}
  \def\iMOSC{$^{(18)}$}
  \def\iNAG{$^{(19)}$}
  \def\iOREG{$^{(20)}$}
  \def\iPAD{$^{(21)}$}
  \def\iPERU{$^{(22)}$}
  \def\iPISA{$^{(23)}$}
  \def\iRUT{$^{(24)}$}
  \def\iRAL{$^{(25)}$}
  \def\iSOGANG{$^{(26)}$}
  \def\iSLAC{$^{(27)}$}
  \def\iTENN{$^{(28)}$}
  \def\iTOH{$^{(29)}$}
  \def\iVAND{$^{(30)}$}
  \def\iWASH{$^{(31)}$}
  \def\iWISC{$^{(32)}$}
  \def\iYALE{$^{(33)}$}
  \def\dead{$^{\dag}$}
  \def\andgen{$^{(a)}$}
  \def\andper{$^{(b)}$}
%
%
$^*$
\mbox{K. Abe                 \unskip,\iNAG}
\mbox{K. Abe                 \unskip,\iTOH}
\mbox{I. Abt                 \unskip,\iILL}
\mbox{T. Akagi               \unskip,\iSLAC}
\mbox{N.J. Allen             \unskip,\iBRUN}
\mbox{W.W. Ash               \unskip,\iSLAC$^\dagger$}
\mbox{D. Aston               \unskip,\iSLAC}
\mbox{K.G. Baird             \unskip,\iRUT}
\mbox{C. Baltay              \unskip,\iYALE}
\mbox{H.R. Band              \unskip,\iWISC
% ...
}
\it
  \vskip \baselineskip                   
  \vskip \baselineskip                   
%
%
%
  \iADEL
     Adelphi University,
     Garden City, New York 11530 \break
  \iBOL
     INFN Sezione di Bologna,
     I-40126 Bologna, Italy \break
  \iBU
     Boston University,
     Boston, Massachusetts 02215 \break
  \iBRUN
     Brunel University,
     Uxbridge, Middlesex UB8 3PH, United Kingdom \break
  \iUCSB
     University of California at Santa Barbara,
     Santa Barbara, California 93106 \break
  \iUCSC
     University of California at Santa Cruz,
     Santa Cruz, California 95064 \break
  \iCIN
     University of Cincinnati,
     Cincinnati, Ohio 45221 \break
  \iCSU
     Colorado State University,
     Fort Collins, Colorado 80523 \break
  \iCOLO
     University of Colorado,
     Boulder, Colorado 80309 \break
  \iCOL
     Columbia University,
     New York, New York 10027 \break
  \iFER
     INFN Sezione di Ferrara and Universit\`a di Ferrara,
     I-44100 Ferrara, Italy \break
  \iFRA
     INFN  Lab. Nazionali di Frascati,
     I-00044 Frascati, Italy \break
  \iILL
     University of Illinois,
     Urbana, Illinois 61801 \break
  \iLBL
     Lawrence Berkeley Laboratory, University of California,
     Berkeley, California 94720 \break
  \iMIT
     Massachusetts Institute of Technology,
     Cambridge, Massachusetts 02139 \break
  \iMASS
     University of Massachusetts,
     Amherst, Massachusetts 01003 \break
  \iMISS
     University of Mississippi,
     University, Mississippi  38677 \break
  \iMOSC
     Moscow State University,
     Institute of Nuclear Physics,
     119899 Moscow,
     Russia    \break
  \iNAG
     Nagoya University,
     Chikusa-ku, Nagoya 464 Japan  \break
  \iOREG
     University of Oregon,
     Eugene, Oregon 97403 \break
  \iPAD
     INFN Sezione di Padova and Universit\`a di Padova,
     I-35100 Padova, Italy \break
  \iPERU
     INFN Sezione di Perugia and Universit\`a di Perugia,
     I-06100 Perugia, Italy \break
  \iPISA
     INFN Sezione di Pisa and Universit\`a di Pisa,
     I-56100 Pisa, Italy \break
  \iRUT
     Rutgers University,
     Piscataway, New Jersey 08855 \break
  \iRAL
     Rutherford Appleton Laboratory,
     Chilton, Didcot, Oxon OX11 0QX United Kingdom \break
  \iSOGANG
     Sogang University,
     Seoul, Korea \break
  \iSLAC
     Stanford Linear Accelerator Center, Stanford University,
     Stanford, California 94309 \break
  \iTENN
     University of Tennessee,
     Knoxville, Tennessee 37996 \break
  \iTOH
     Tohoku University,
     Sendai 980 Japan \break
  \iVAND
     Vanderbilt University,
     Nashville, Tennessee 37235 \break
  \iWASH
     University of Washington,
     Seattle, Washington 98195 \break
  \iWISC
     University of Wisconsin,
     Madison, Wisconsin 53706 \break
  \iYALE
     Yale University,
     New Haven, Connecticut 06511 \break
  \dead
     Deceased \break
  \andgen
     Also at the Universit\`a di Genova \break
  \andper
     Also at the Universit\`a di Perugia \break
\rm
%

\end{center}

\narrowtext
The SLD Collaboration has performed measurements of the left-right 
cross section asymmetry A$_{LR} = (\xl-\xr)/(\xl+\xr)$ in the production of
\z0 bosons by \ee collisions \cite{SLDALR1,SLDALR2,SLDALR3}.  In the Standard 
Model of the electroweak interactions, to first order, this gives the electron 
left-right asymmetry factor $A_e = 2v_e a_e/(v^2_e + a^2_e)$ from \cite{FT1}
\begin{equation}
\alrraw\ =\ \absp \alr\ =\ \absp ~A_e \label{alrraw}
\end{equation}
where $\pole$ is the electron-beam longitudinal polarization, and $v_e$ and
$a_e$ are the vector and axial vector coupling constants between the Z$^0$ and
electron.  
The forward-backward fermion asymmetries in \z0
decays can also be used to provide independent information on the electron
couplings to the \z0.
The forward-backward fermion asymmetry 
at the \z0 pole (excluding \ee final states) is given by
\begin{equation}
A_{FB,f}(\pole) = -g(a){\pole -A_e\over 1-\pole A_e} ~A_f \label{afbp}
\end{equation}
where $g(a)\!=\!a/(1+{1\over 3}a^2)$, $0\!<\!a\!\leq\!1$, 
$a\!=\!\vert\cos\theta\vert_{max}$, $\cos\theta$ describes the angle between 
the outgoing fermion $f$ and the direction of the incident electron, $max$ 
refers to the  aperture limit of the detector, and
$A_f = 2v_f a_f/(v^2_f + a^2_f)$.  We can define 
$A^L_{FB,f}\equiv A_{FB,f}(-\absp)$ and $A^R_{FB,f}\equiv A_{FB,f}(\absp)$ 
as the forward-backward asymmetries for events produced with left and 
right-handed beam polarization respectively.

These asymmetries can be related to observable charge asymmetries
\cite{ALEPHQFB,DELPHIQFB,OPALQFB}.  At the parton level the fermion 
asymmetries for a quark anti-quark final state give the following average 
charges in the forward and backward hemispheres of left-handed events:
\begin{eqnarray}
<\qr^L_{F,f}> & = & ~~q_f\ A^L_{FB,f}
\nonumber \\
<\qr^L_{B,f}> & = & -q_f\ A^L_{FB,f}
\label{AtoQ}
\end{eqnarray}
where $q_f$ is the charge of the outgoing fermion. Similar expressions 
hold for right-handed events. These average charges
can then be combined into the forward-backward charge flows, or asymmetries.  
For left-handed events:
\begin{eqnarray}
  <\qr^L_{FB,f}> & \equiv & <\qr^L_{F,f}> - <\qr^L_{B,f}>
                 ~=~ 2\ q_f\ A^L_{FB,f}
\label{QFBLRf}
\end{eqnarray}
with a similar expression for right-handed events. 

The flavor-inclusive observables for the polarized $<\tqr_{FB}>$ and 
unpolarized $<Q_{FB}>$ forward-backward charge flows, which are measured at the
final state hadron level, can be defined by summing over all flavors, weighting 
by the flavor production rate, and including dilution factors 
$0~<~d_f~<~1$ to account for a reduction in the measured charge magnitudes due 
to QCD corrections, hadronization effects, and $B\bar B$ mixing \cite{FT2} as
follows:
\begin{eqnarray}
<\tqr_{FB}> &\equiv & <\qr^L_{FB}> f_L\ - <\qr^R_{FB}> f_R 
            ~=~  2\ g(a)\ \absp \ \sum_f d_f q_f R_f A_f
\label{QFBPOL}
\end{eqnarray}
\begin{eqnarray}
<Q_{FB}> &\equiv & <\qr^L_{FB}> f_L\ + <\qr^R_{FB}> f_R
         ~=~ 2\ g(a)\ A_e \ \sum_f d_f q_f R_f A_f
\label{QFB}
\end{eqnarray}
where $f_L={1\over2}(1+\absp A_e)$ and $f_R={1\over2}(1-\absp A_e)$ are the
fractions of left- and right-handed events, $R_f=\Gamma_f/\Gamma_{had}$, 
$\Gamma_f$ is the partial width for the decay $Z\to f\bar f$, and
$\Gamma_{had}$ is the total hadronic width of the Z. The
quantities $<Q^L_{FB}>$ and $<Q^R_{FB}>$ represent the mean, 
flavor-inclusive, forward-backward charge flows in left- and right-handed 
events. These quantities are measured using the momentum-weighted charge
technique described below. 

The ratio of these charge asymmetries has the simple form
\begin{eqnarray}
\qlrraw & \equiv & {<\qr_{FB}> \over <\tqr_{FB}>}
           ~=~    {A_e \over \absp}.
\label{QLR}
\end{eqnarray}

The expression for $\qlrraw$ shows that uncertainties in the detector 
acceptance, charge measurement, and the dilution factors cancel out, thus 
effectively eliminating the
dependence on Monte Carlo simulation for such corrections. Many systematic 
instrumental effects were investigated and are discussed below.

By measuring the quantity $\qlrraw\absp$, ~$A_e$ can be obtained in a manner 
largely independent of the $A^{obs}_{LR}$ measurement \cite{FT3}.
Furthermore, the two measurements can be combined to yield $A_e$ without a 
measurement of the electron polarization, using the expression
\begin{equation}
A_e\ =\ \sqrt{\alrraw \times \qlrraw}.
\label{AeNOP}
\end{equation}
This determination of A$_e$ is not independent of the more precise
measurement using $A^{obs}_{LR}$ and the longitudinal polarization that has 
been published elsewhere \cite{SLDALR2,SLDALR3}. 

In this paper, we present the first measurement of $A_e$ from 
$\qlrraw$ and the electron-beam polarization.  We also
present an alternative measurement of $A_e$ from $\alrraw$ and $\qlrraw$ that 
does not require knowledge of the polarization magnitude.

Details of the SLAC Linear Collider (SLC), the polarized electron source, the 
measurement of the electron-beam polarization with the Compton polarimeter, 
and the SLD have been given elsewhere \cite{SLDALR1,SLDALR2,POL}.
The results presented in this article are based upon a sample of 
data corresponding to an integrated luminosity of $5.1~pb^{-1}$.
The data were recorded at a mean center-of-mass energy of $91.27\pm 0.02$ GeV
during the 1993 and 1994-1995 runs of the SLC.

The momenta of charged particles were measured in the
central drift chamber (CDC). Accepted particles were required 
to have: (i) a minimum momentum transverse to the beam axis $>$ 0.15 GeV/c;
(ii) a polar angle $\theta$ with respect to the beam axis satisfying 
$\vert\cos\theta\vert < 0.8$; and (iii) a point of closest approach to the 
beam axis within a cylinder of $5~cm$ radius and $10~cm$ half-length about 
the interaction point.  If any remaining particle in an event had a total 
momentum $>$ 55 GeV/c the event was rejected.

Each event was divided into two hemispheres by a plane transverse to the 
thrust axis \cite{THRUST} which was determined using all accepted charged 
particles in the event. Hadronic events were selected by the following 
requirements:
(i) the polar angle of the thrust axis satisfied 
$\vert\cos\theta_{T}\vert < 0.7$;  (ii) there were at least three particles per 
hemisphere; (iii) the total energy of the particles in the event (assuming the 
particles to be pions) was greater than 20\% of the center-of-mass energy;
(iv) the scalar sum per hemisphere of particle momentum components parallel to 
the thrust axis was greater than 10\% of the beam energy; and (v) the 
invariant mass of the particles in at least one hemisphere was greater than 
$2~GeV/c^2$.  A total of 49,850 hadronic Z decays produced by left-handed 
electrons  and 39,988 produced by right-handed electrons
were obtained with an estimated non-Z  background of less than 
.05\% \cite{FT4}. The effect of the residual $\tau^+\tau^-$ events on the value
of A$_Q^{obs}$ was estimated to be (0.028 $\pm$ 0.012)\% which is negligible. 
This is relevant because final-state polarization effects in this channel 
complicate its contribution to this quantity. The luminosity-weighted 
polarization for this sample of events was $0.730\pm 0.008$, where the 
error is predominantly systematic \cite{FT5}. 

The forward-backward charge asymmetries were determined in the following 
manner.  A unit vector along the thrust axis,~$\bf{\hat T}$, ~was chosen such 
that ${\bf{\hat T}\cdot} {\bf p}_{e^-} > 0$, where ${\bf p}_{e^-}$ is the
electron beam direction.  Tracks with momentum vector {\bf p} were defined as 
forward 
if ${\bf p \cdot {\hat T}} > 0$, ~and backward otherwise.  The weighted 
charge in the forward hemisphere was then calculated for each event from
\begin{equation}
Q_F = {\sum_{{\bf p}_i \cdot {\bf\hat T}>0}
       \vert {\bf p}_i \cdot {\bf\hat T}\vert q_i \over
       \sum_{{\bf p}_i \cdot {\bf\hat T}>0}
       \vert {\bf p}_i \cdot {\bf\hat T}\vert}
\label{QF}
\end{equation}
where $q_i$ is the charge of particle $i$. The charge in the backward
hemisphere, $Q_B$, was determined in a similar manner for tracks with 
${\bf p \cdot {\hat T}} < 0$.  The quantity $Q_{FB}=Q_F-Q_B$ was then found for
each event.  

\begin{figure}
\centerline{\epsfig{file=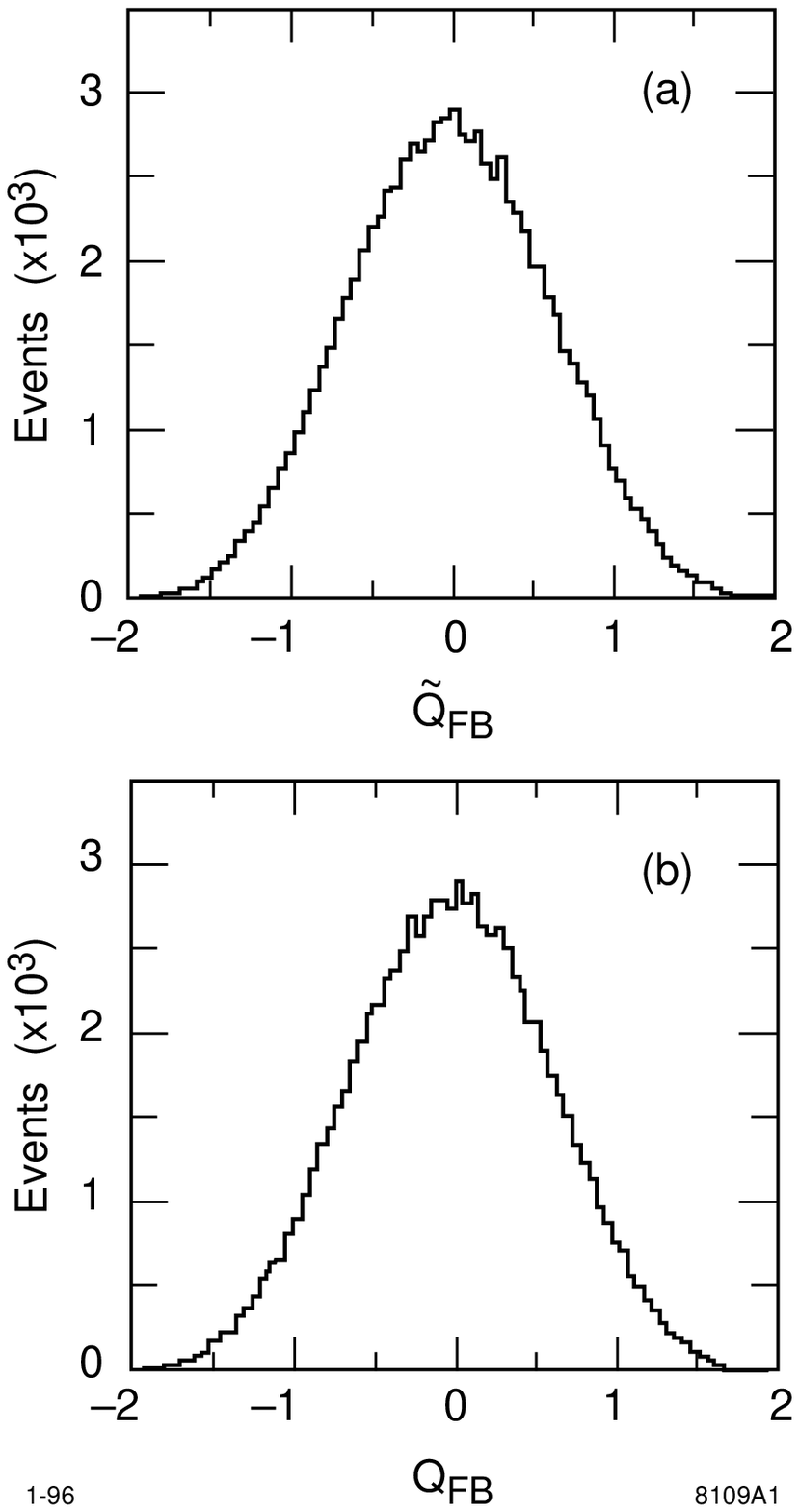,width=3.35 in}}
\caption{Distributions of the polarized (a) and unpolarized (b), 
forward-backward charge flows.}
\label{qfbfigs}
\end{figure}

The distribution of $Q_{FB}$ was formed separately for left- and right-handed
events. The distributions for $<\!\tqr_{FB}\!>$ and $<\!Q_{FB}\!>$ were
obtained in accordance with \mbox{Eqs.~(\ref{QFBPOL})~and~(\ref{QFB})}
and are shown in Fig.~\ref{qfbfigs}.  
The averages  $<\!\tqr_{FB}\!>$ and $<\!Q_{FB}\!>$ were 
obtained from their corresponding distributions \cite {FT6}.  
Then $\qlrraw$ was determined using \mbox{Eq.~(\ref{QLR})}.  A value for 
$A^{obs}_{LR}$ was also obtained using the number of accepted left- and 
right-handed events.  These results are summarized in 
\mbox{Table~\ref{results}}.

We investigated a number of possible systematic errors due to biases in
instrumentation, analysis misindentification, charge dependent nuclear 
interactions of low momentum hadrons, unphysical measured momenta, 
material asymmetries, and various backgrounds. We studied the possibility of a
charge-dependent, forward-backward bias in the measured track sagitta, 
or momenta, by  means of the dimuon and Bhabha events in the data sample. 
This can produce an artificial  change in $<Q_{FB}>$, while affecting 
$<\tqr_{FB}>$ very little, thus biasing $A^{obs}_Q$ \cite{QBIAS}.
This study led to a $(-1.6\pm 6.5)\%$ change in $A^{obs}_Q$. This error was the
largest of the systematic errors studied. The systematic errors on
the value of A$_Q^{obs}$ resulting from these studies are presented in Table II.
The value for the left-right charge asymmetry, before radiative 
corrections and including the systematic error from Table II, is
\begin{equation}
 A^{obs}_Q = 0.225\pm 0.056 \; (stat.)\pm 0.019  \; (syst.).
\label{QLRINV}
\end{equation}
To obtain the relevant quantities A$_e$ and $\swein$ from A$_Q^{obs}$ we must
correct Eqs. 7 for Z-$\gamma$ interference, $\gamma$ exchange and radiative 
corrections. These were made to the measured asymmetries using the ZFITTER
program \cite{ZFITTER}. The cancellation of the flavor sum in Eq.~\ref{QLR} 
is not preserved by these higher order processes, and Eqs.~\ref{QFBPOL} 
and~\ref{QFB} must be used with ZFITTER to obtain $A_e/|P_e|$.
The charge dilution factors $d_f$ were varied by $\pm 20\%$ in a manner
that maximizes the variation of the radiative correction to $\qlr$.  This
results in an uncertainty of $\pm 4\%$ in the corrected value of $\qlr$.
After these corrections, the following 
values are obtained:
\begin{eqnarray}
A_e    & = & 0.162 \pm 0.041 \; (stat.)\pm 0.014 \; (syst.) \nonumber\\
\swein & = & 0.2297 \pm 0.0052 \; (stat.)\pm 0.0018 \; (syst.).
\end{eqnarray}
These results are largely independent of those previously obtained by SLD 
from $\alr$, and are in good agreement with them. 

We can also obtain $A_e$ from $\qlrraw$ and $\alrraw$ using 
\mbox{Eq.~(\ref{AeNOP})},
without the use of the Compton-measured polarization.  After radiative 
corrections to the measured results, we obtain:
\begin{eqnarray}
A_e    & = & 0.1574 \pm 0.0197 \; (stat.) \pm 0.0067 \; (syst.) \nonumber\\
\swein & = & 0.2302 \pm 0.0025 \; (stat.) \pm 0.0009 \; (syst.),\nonumber
\end{eqnarray}
This result is not independent of those
obtained from $A_{LR}$ and $A^{obs}_Q$ separately.  Rather, it is an
alternative measurement of $A_e$ and $\swein$ that does not use the 
measured polarization. This is a completely new technique in the determination
of these quantities. 
%
These results can be compared with the latest value of $\swein$ = 0.23049
$\pm$ 0.00050, obtained
directly from a measurement of $A_{LR}$ and the electron longitudinal
polarization \cite{SLDALR3}. 

We thank the personnel of the SLAC accelerator department and the technical
staffs of our collaborating institutions for their efforts which resulted in
the successful operation of the SLC and the SLD. This work was supported by
the Department of Energy,; The National Science Foundation; the Instituto
Nazionale di Fisica Nucleare of Italy; the Japan-US Cooperative Research
Project on High Energy Physics; and the Science and Engineering Research
Council of the United Kingdom.
%

%
%
%
%
\begin{table}
\caption{Summary of Results}
\label{results}
\begin{tabular}{lccc}
Quantity                                          &     Value\\
\hline
$f_L, f_R$                                        & 0.5549, 0.4451\\
$<\!{Q}^L_{FB}\!>$                                & $      -0.0408\pm 0.0027$\\
$<\!{Q}^R_{FB}\!>$                                & $\;\;\: 0.0322\pm 0.0031$\\
$<\!\tilde{Q}_{FB}\!>$                            & $     -0.03697\pm 0.00204$\\
$<\!Q_{FB}\!>$                                    & $     -0.00831\pm 0.00204$\\
$A_Q^{obs}      $                                 & $\;\;\: 0.2247\pm 0.0556$\\
$A_{LR}^{obs}$                                    & $\;\;\: 0.1098\pm 0.0033$\\
\end{tabular}
\end{table}
\begin{table}
\caption{Summary of Systematic Errors}
\label{sys}
\begin{tabular}{lc}
                      & $\delta A^{obs}_Q/A^{obs}_Q$ \\
Source of uncertainty &           (\%) \\
\hline
q dependent, F-B sagitta bias                                  & 6.5 \\
q independent, F-B sagitta bias                                & 0.5 \\
q independent, F-B track efficiency biases                     & 0.2 \\
unphysical $p_{tot}$ tracks                                    & 3.3 \\
F-B asymmetry of SLD central material                          & 1.5 \\
$e^+e^-$ final state backgrounds                               & 0.5 \\
two photon backgrounds                                         & 0.7 \\
radiative corrections                                          & 4.0 \\
polarization measurement (for result~(\ref{QLRINV}) only)      & 1.1 \\
Residual $\tau^+\tau^-$ effect                                 & 0.03 \\
SLC track backgrounds                                          & 0.02 \\
\hline
Total                                                          & 8.7 \\
\end{tabular}
\end{table}

\begin{references}
\bibitem{FT1} In Ref. 1, 2 we use A$_m$ instead of A$_{LR}^{obs}$.
\bibitem{FT2} It is assumed that the difference in the $d_f$ factors for 
processes initiated by left handed and right handed electrons is negligible.
\bibitem{FT3} The correlation 
between $\alrraw$ and $\qlrraw$ is approximately -6\%.  Likewise, the 
correlation coefficient between $A_{LR}$ and $Q_{LR}$ is similar in value
for the measurement presented in this letter, since the error in 
the measured polarization is small compared to the statistical errors 
in $\alrraw$ and $\qlrraw$.
\bibitem{FT4} Our background is 
less than in previous studies of the SLD collaboration due to the more 
stringent cuts in our hadronic sample \cite{QCDPUB}.
\bibitem{FT5} This is the luminosity-weighted average polarization from two
runs with polarization of 0.63 and 0.78. \cite{SLDALR2,POL}.
\bibitem{FT6} In computing the charge asymmetries for each event, a correction 
for a charge-dependent, forward-backward bias to the track sagitta was applied 
to each track's momentum as determined by our systematic error studies.
\bibitem{QBIAS}This is because in $<\tilde{Q}_{FB}>$ we measure the 
difference of the charge flow for left- and right-handed electrons, canceling
these types of biases, while in $<Q_{FB}>$ we measure the sum.
\end{references}
\end{document}